\documentclass[prb,showpacs,twocolumn,amsmath,amssymb]{revtex4}

\usepackage{graphicx}
\usepackage{dcolumn}
\usepackage{bm}

\newcommand{\Vvec}[1]{\mbox{\boldmath$#1$}}

\begin{document}


\title{Superconductivity from a long-range interaction:\\
a crossover between the electron gas and the lattice model}

\author{Seiichiro Onari}
\altaffiliation[Present address: ]{Graduate School of Engineering, Nagoya University, Chikusa, Nagoya 464-8603, Japan.}
\author{Ryotaro Arita$^1$}
\altaffiliation[On leave of absence from ]{Department of Physics,
University of Tokyo.}
\author{Kazuhiko Kuroki$^2$}
\author{Hideo Aoki}
\affiliation{Department of Physics, University of Tokyo, Hongo,
Tokyo 113-0033, Japan}
\affiliation{$^1$Max Planck Institute for Solid State Research,
Heisenbergstr. 1, Stuttgart 70569, Germany}
\affiliation{$^2$Department of Applied Physics and Chemistry,
University of Electro-Communications, Chofu, Tokyo 182-8585, Japan}

\date{\today}

\begin{abstract}
We explore how the 
superconductivity arising from the on-site electron-electron
repulsion will change when the 
repulsion is changed to a long-ranged, $1/r$-like one 
by introducing an extended Hubbard model with 
the repulsion extending to distant (12th) neighbors.  
With a simplified fluctuation-exchange 
approximation, we have found for the square lattice that 
(i) as the band filling becomes dilute enough, 
the charge susceptibility becomes comparable with the 
spin susceptibility, where $p$ and then $s$ pairings become 
dominant, in agreement with the result for the electron gas by 
Takada, while (ii) the $d$-wave, which reflects the lattice 
structure, dominates well away from the half filling.  
All these can be understood in terms of the spin and
 charge structures along with the shape and size of the Fermi surface.
\end{abstract}


\maketitle


While the discovery of high-Tc cuprates has kicked off intensive 
studies of electron mechanisms of superconductivity, 
these have been primarily focused on the pairing 
from {\it short-ranged} electron-electron repulsion.  
This is a reasonable assumption for transition metal 
compounds, since the main interaction between the $d$ electrons should be 
short-ranged, as captured in the Hubbard model with 
an on-site repulsion.  
However, if we go back to the history of electron mechanisms 
of superconductivity, there is an important predecessor --- 
the electron gas with the long-range Coulomb interaction, 
one of the most fundamental problems in the condensed-matter 
physics.  The problem has a long history, where 
Kohn and Luttinger\cite{Kohn}, as early as in the 1960's, have 
suggested that normal states in the electron gas 
should become unstable in favor of a superconducting state, 
which has later been proved when the gas is dilute enough\cite{chubukov}.
There are two important factors that discriminate the electron gas 
from electron systems with short-range 
interactions.  One is that the gas becomes more strongly interacting 
(i.e., larger ratio of the interaction to the kinetic energy) 
for the more dilute concentrations.  More importantly, the long-range 
interaction can make the {\it charge fluctuations} as large as the spin 
fluctuations as contrasted with systems with short-range 
interactions where the spin fluctuations dominate, so that 
the fluctuation-mediated interaction should change with the range of the
interaction.  

A fascinating question then is: 
what will become of the superconductivity in the electron 
system with short-range repulsions when the interaction range 
is increased to approach $1/r$.  Besides the problem of the 
interaction range, there is another essential question: 
if we consider a lattice systems as in the 
Hubbard model, the band filling is a crucial parameter 
with the half-filling being a special point, 
which controls Mott's metal-insulator transition as well as 
the Fermi surface nesting, which in turn dominates the 
spin fluctuation.  
Indeed, we have a $d$-wave pairing mediated by spin fluctuations 
in the on-site Hubbard model around the half filling, as 
theoretically suggested with, among other methods, the quantum Monte Carlo 
method\cite{kuroki0}, the fluctuation exchange (FLEX) 
approximation\cite{bickers1,bickers2,dahm,bennemann}, and the dynamical
cluster approximation\cite{maier}.  
By contrast the electron gas in a continuous space, 
only characterized by the electron concentration 
(or, more precisely, $r_s$, the mean electron separation 
measured by the Bohr radius) with a circular Fermi surface, 
has no such special fillings.  
Takada\cite{takada2} has intensively studied 
superconductivity in the three-dimensional electron gas, and has concluded 
that a $p$-wave pairing should occur for $r_s > 3.3$, which 
gives way to an $s$-wave pairing when the gas becomes 
more dilute ($r_s > 8.6$), where the pairing is interpreted 
to arise from charge 
fluctuations including plasmons\cite{takada3}.  

One step toward the increased range of 
interaction is to consider the extended Hubbard model which 
takes account of the nearest-neighbor repulsion $V$. 
The extended Hubbard model has been
studied with various theoretical 
methods\cite{Zhang,Tesanovic,Murakami,Seo,Onozawa,Scalapino3D,kobayashi,tanaka,esirgen3,arita2,onari-cdw,Merino,Sano,Kuroki-Kusakabe}.  
For instance, Refs.\onlinecite{arita2,onari-cdw} address, with FLEX, the 
question of how the 
enhanced charge fluctuations in the presence of $V$ should affect 
the pairing symmetry, where the result indicates that 
a triplet pairing as well as singlet ones appear in the phase diagram, 
whose mechanism can be traced back 
to the structure in the charge and spin susceptibilities.

Given this background, 
the purpose of the present work is to explore what happens to 
the pairing symmetry as we make 
the interaction more long-ranged and closer to the Coulombic interaction. 
For this purpose we take a model where 
the interaction is extended to distant (12th here) neighbors 
for the  square lattice, which is studied with a simplified FLEX.  
We then question: (i) to what extent 
the effects of the lattice persist as we go away from the half filling, 
and (ii) how the pairing crosses over to the pairing in the 
electron gas in the dilute regime.   
We shall show that (i) the pairing symmetries ($d_{x^2-y^2}$,
$d_{xy}$) reflecting 
lattice structure persist well away ($n \gtrsim 0.2$) from the half filled 
band, and (ii) in the dilute ($n \lesssim 0.2$) regime 
the pairing symmetries are $s$ and $p$ in agreement with those for 
the electron gas.  
The key factors controlling the dominant symmetries 
are identified to be the structure (peak positions, intensities and widths) 
of the charge and spin susceptibilities 
(along with the size of the Fermi surface in the dilute regime). 

\begin{figure}[htdp]
\begin{center}
\includegraphics[height=35mm]{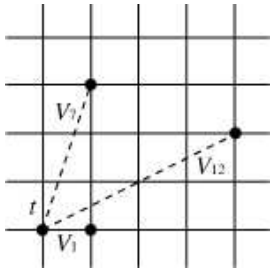}
\caption{A square lattice with the long-range repulsion $V_n$.}
\label{longrange}
\end{center}
\end{figure}

We first introduce an extended Hubbard model 
with long-range interactions (Fig. \ref{longrange}),
{\small
\begin{eqnarray}
{\cal H} = -t\sum_{ij}^{\rm nn}\sum_{\sigma}c_{i\sigma}^{\dagger}c_{j\sigma}+U\sum_{i}n_{i\uparrow}n_{i\downarrow}+\frac{1}{2}\sum_{ij}\sum_{\sigma\sigma'}V_{ij}n_{i\sigma}n_{j\sigma'}.
\end{eqnarray}
}
\noindent 
Here $t(=1)$ is the nearest-neighbor transfer, taken to be 
the unit of energy, and 
the off-site Coulomb repulsion $V_{ij}$ extends 
up to the 12th neighbor for the square lattice as depicted 
in Fig.\ref{longrange}, where 
the magnitude of the $n$th-neighbor repulsion 
$V_n=V_{|i-j|}$ $(V_1\equiv V)$ is taken to be inversely proportional to the 
distance (e.g., $V_{12}=V/\sqrt{20}$). 

We have previously extended\cite{onari-cdw} the FLEX formalism to the 
extended Hubbard model with the nearest-neighbor interaction, but it 
is difficult to extend this beyond the second neighbors.  
So we employ here a simplified FLEX, where 
full bubble diagrams and restricted ladder diagrams 
(that include only the $U$ 
term) are considered for the effective interaction 
in a self-consistent manner. 
So this is a kind of RPA with all the Green's functions dressed, 
which was adopted 
in Refs.\cite{kobayashi,tanaka} for the extended Hubbard model 
with the nearest-neighbor interaction.  
The simplified FLEX still 
belongs to ``conserving approximations'' formulated by Baym and
Kadanoff\cite{baym-kadanoff,baym}.  
The FLEX neglects vertex corrections, but 
it has been known that the correction can lead to some problems
in self-consistent calculations, especially in the dilute regime.\cite{Holm}
So here we opt for FLEX and assume that the vertex correction
would not significantly affect the competition between
different pairing symmetries (although $T_c$, not discussed here,
may be affected).

The spin $(\chi_{\rm sp})$ and charge $(\chi_{\rm ch})$ susceptibilities
are then 
$\chi_{\rm sp}(q)=\frac{\overline{\chi}(q)}{1-U\overline{\chi}(q)}$, 
$\chi_{\rm ch}(q)=\frac{\overline{\chi}(q)}{1+[U+2V(q)]\overline{\chi}(q)}$, 
where the irreducible susceptibility is given by
$\overline{\chi}(q)=-(T/N)\sum_kG(k+q)G(k)$.  
Here $q\equiv({\Vvec q},\epsilon_n)$ with $\epsilon_n\equiv2n\pi T$ being the
Matsubara frequencies for bosons and $k=({\Vvec k}, \omega_n)$ with
$\omega_n=(2n-1)\pi T$ for fermions, 
and 
$V(q)=2V_1(\cos q_x+\cos q_y)+4V_2(\cos q_x\cos q_y)
+2V_3(\cos2q_x+\cos2q_y)+\cdots$.
The self-energy is given as
{\scriptsize
\begin{eqnarray}
&\Sigma(k)&=\frac{T}{N}\sum_{k'}\left\{-V(k-k')+\frac{3}{2}U^2\chi_{\rm sp}(k-k')\right.\nonumber\\
&+&\!\!\!\!\left.\left[\frac{1}{2}U^2+2UV(k-k')+2V^2(k-k')\right]\chi_{\rm ch}(k-k')\right\}G(k').
\end{eqnarray}
}
The gap function and $T_c$ are obtained with \'Eliashberg's equation,
\begin{eqnarray}
\lambda\phi(k)=-\frac{T}{N}\sum_{k'}\Gamma(k-k')G(k')G(-k')\phi(k'),
\end{eqnarray}
where $\lambda=1$ corresponds to $T=T_c$, and the pairing interactions 
$\Gamma_s (\Gamma_t)$ for the singlet (triplet) channels are
{\scriptsize
\begin{eqnarray}
\Gamma_s(q)\!\!\!&=&\!\!\!U+V(q)+\frac{3}{2}U^2\chi_{\rm sp}-\left[\frac{1}{2}U^2+2UV(q)+2V^2(q)\right]\!\!\chi_{\rm ch}(q)\label{rpas},\\
\Gamma_t(q)\!\!\!&=&\!\!\!V(q)-\frac{1}{2}U^2\chi_{\rm sp}-\left[\frac{1}{2}U^2+2UV(q)+2V^2(q)\right]\!\!\chi_{\rm ch}(q)\label{rpat}.
\end{eqnarray}
}
\noindent Here we take $N=32\times32$ ${\Vvec k}$-point meshes and the
Matsubara frequencies $\omega_n$ from $-(2N_c-1)\pi T$ to $(2N_c-1)\pi
T$ with $N_c=16384$, $T=0.01$, $U=4.0$.

\begin{figure}[htdp]
\begin{center}
\includegraphics[height=100mm]{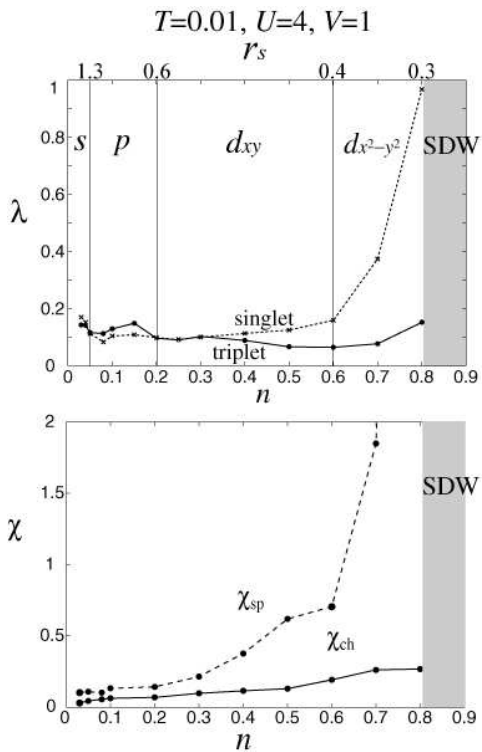}
\caption{Top: Maximum eigenvalue, $\lambda$, of \'Eliashberg's equation 
in the triplet (solid line) and singlet (dotted) channels 
with the dominant symmetry indicated along with the boundary values 
of $r_s$ on the top axis.  The SDW phase is
identified from the divergence of the spin susceptibility. 
Bottom: Charge 
(solid line) and spin (dotted) susceptibilities against 
$n$. 
}
\label{rpaelia}
\end{center}
\end{figure}

\begin{figure}[htdp]
\begin{center}
\includegraphics[height=60mm]{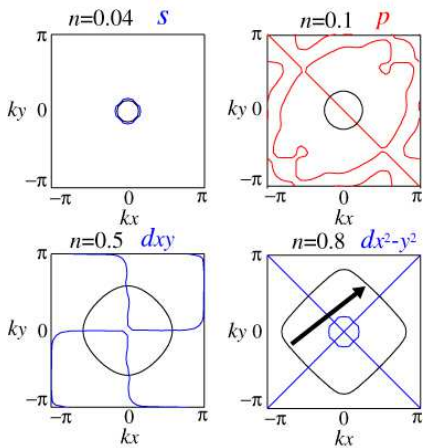}
\caption{Fermi surface (obtained as $\epsilon^0_{\Vvec k}+
{\rm Re}\Sigma({\Vvec k})
=\mu$; black line) and nodes in the dominant gap function 
(blue lines for singlet and red for triplet) for 
$0.04\leq n \leq 0.8$ with $U=4.0$ and $V=1.0$. The arrow indicates the main 
scattering process mediated by spin fluctuations.
}
\label{rpagap}
\end{center}
\end{figure}

We first show the eigenvalue, $\lambda$, of \'Eliashberg's equation, 
along with the 
charge $(\chi_{\rm ch})$ and spin $(\chi_{\rm sp})$ susceptibilities against 
the band filling $n$ for $V=1.0$ in Fig.\ref{rpaelia}. While $\lambda$
for $T=0.01$ is
still much smaller than unity, it is difficult to go down to lower
temperatures, given a huge computational demand for the long-range 
model. So we assume that the dominant pairing
symmetry is the one that has the largest $\lambda$ at $T=0.01$ which is
higher than $T_c$.
This amounts to assuming that, while the
position of boundary may shift as $T\rightarrow0$, the order in
which dominant symmetry changes is not altered.  
We see that the dominant pairing is $d_{x^2-y^2}\rightarrow
d_{xy}\rightarrow p\rightarrow s$ 
for the band filling is decreased from the half-filling.
Figure \ref{rpagap} depicts the four gap functions in ${\Vvec k}$ space 
for typical values of $n$.  

\begin{figure}[htdp]
\begin{center}
\includegraphics[height=40mm]{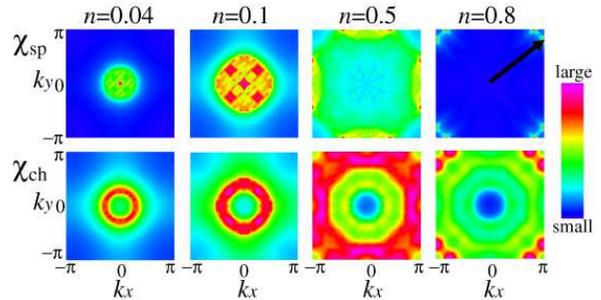}
\caption{Spin susceptibility $\chi_{\rm sp}$ (top panels) 
and charge susceptibility $\chi_{\rm ch}$ (bottom) 
in ${\Vvec k}$ space 
for typical values of the band filling $n$. 
To make the peaks clearer, the 
color coding differs from frame to frame.
}
\label{stsus}
\end{center}
\end{figure}

To explore why the dominant gap symmetry changes in such a way, 
we have plotted the charge and spin susceptibilities in 
${\Vvec k}$ space in Fig. \ref{stsus}.  
Let us start with the region $0.8 \gtrsim n \gtrsim 0.3$, where 
the symmetry changes from $d_{x^2-y^2}$ to $d_{xy}$ as $n$ is decreased.  
For $n=0.8$ the peak positions in the spin susceptibility are situated around 
${\Vvec k}=(\pi,\pi)$ as marked with an arrow in Fig. \ref{stsus}, 
which accounts for the $d_{x^2-y^2}$ pairing.  
For $n=0.5$, the peaks shift to 
$(0,\pm\pi), (\pm\pi,0)$, which should be 
why the $d_{xy}$ gap function, which changes sign 
across the pair hopping $(0,\pm\pi), (\pm\pi,0)$, takes over. 

If we further decrease the band filling, we see from Fig.\ref{rpaelia}
that $\chi_{\rm ch}$, which is originally much smaller than 
$\chi_{\rm sp}$ for $n>0.3$, becomes comparable with $\chi_{\rm sp}$.  
When the spin-fluctuation mediated interaction is 
dominant (as is the case with short-range repulsions), 
the singlet pairing
interaction $\Gamma_s$ (eq.\ref{rpas}) has the
$\chi_{\rm sp}$-term whose coefficient is three times larger 
in magnitude than that for the triplet, $\Gamma_t$ 
(eq.\ref{rpat}).  So 
the singlet superconductivity is generally favored when the interaction 
is short-ranged (i.e., the spin fluctuation is dominant).  
Conversely, if spin and charge susceptibilities are comparable, 
the situation may be inverted, which indeed occurs for 
the triplet $p$-wave around $0.2 \gtrsim n \gtrsim 0.05$. 

\begin{figure}[htdp]
\begin{center}
\includegraphics[height=45mm]{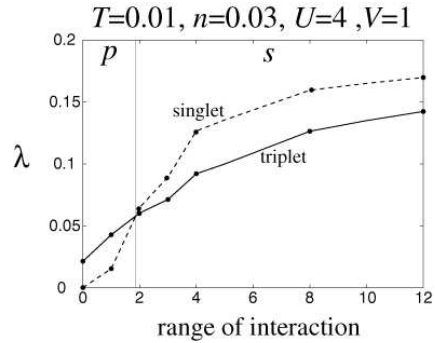}
\caption{The maximum eigenvalue of \'Eliashberg's equation for triplet
 (solid line) and for singlet (dotted) against the range of the
 interaction (in units of the lattice constant) for $n=0.03$ and $V=1.0$.
}
\label{rdep}
\end{center}
\end{figure}

In the most dilute regime ($n \lesssim 0.05$) 
an $s$-wave has the largest $\lambda$.  
Since the ratio $\chi_{\rm ch}/\chi_{\rm sp}$ does not drastically 
change in this region, 
the appearance of $s$ requires an explanation.  
Here we can note another factor that affects the pairing.  
This is a very general observation that 
nodes in the gap function, even when they are 
necessary, act to lower $T_C$, since some of 
the pair scatterings around each node work against the paired state.  
In the present context, this occurs for the $p$ wave for which 
the Fermi surface shrinks with decreasing $n$, so that 
the fraction of the phase space volume for the unfavorable region around 
the nodes increases.  We identify this to be the reason 
why $p$ gives way to $s$ in the most dilute region.  
The competition between $s$ and $p$ when 
the range of interaction is varied is displayed in Fig.\ref{rdep}, 
where we see that 
$s$-wave tends to be favored as the interaction becomes longer-ranged.  
We can understand this in terms of the pairing interaction, since
the spin structure in this regime is peaked at ${\Vvec k}=0$,
which inhibits the singlet $s$ and assists the triplet $p$,
and the peak rapidly decreases as the range is increased.

Let us finally compare the present result with the 
situation in the electron gas.  
Takada\cite{takada2} has obtained, with the Kukkonen-Overhauser 
method\cite{kukkonen}, a phase diagram as a function of the
electronic-density parameter $r_s$ for the electron gas, where  
$p$-wave (at $r_s=3.3$) and then $s$-wave (at $r_s=8.6$) 
appear as $r_s$ is increased (i.e., as the gas becomes more dilute).
To quantify the connection between the electron gas and
the present lattice model (Fig.\ref{rpaelia}), we have to define 
$r_s$ for the latter.  In the effective-mass sense 
we have $r_s=\sqrt{2}/(a_B^*k_F)$, 
where $a_B^*=(4\pi\epsilon_0\hbar^2)/(m^*e^2)$ is the 
effective Bohr radius, and 
$m^*\sim \hbar^2/(2a^2t)$ the effective mass with $a$ being the 
lattice constant.  Still, the definition can be a bit tricky, 
since the interaction is not truly $1/r$.  
Here we eliminate $e$ from the above and 
$V=e^2/(4\pi\epsilon_0a)$ to define $r_s\equiv V/(\sqrt{2}ak_Ft)$, 
where $k_F$ in the radius of the Fermi circle 
when the band is dilute, while 
for $n$ close to half-filling we define $k_F$ as the largest radial 
distance from $\Gamma$ as a measure of the Fermi surface.

\begin{figure}[h]
\begin{center}
\includegraphics[height=45mm]{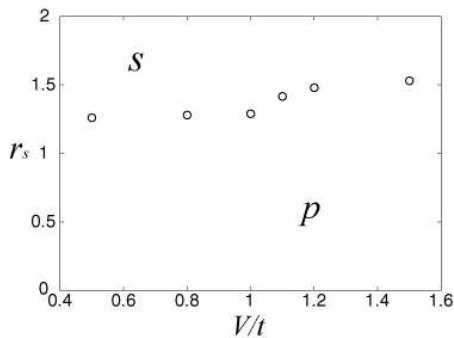}
\caption{The critical value of $r_s$ between $s$-wave and $p$-wave plotted 
against $V/t$ for $U=4.0$ at $T=0.01$.}
\label{rs-sp}
\end{center}
\end{figure}


We first examine how the boundary value of $r_s$ between $s$- and $p$-waves 
changes with $V$ in Fig.\ref{rs-sp}.  We can see that the boundary is 
almost flat, especially so for 
$V/t\leq1$ where the dispersion around small $k_F$ is close to those of 
the electron gas.  So the result is consistent in that the phase 
is basically determined by $r_s$ alone as in the electron gas.

If we turn to the numerical values of $r_s$ at the boundary, 
we see that, although $p$- and $s$-waves 
appear in the dilute concentration regime in both of the 
present and electron-gas\cite{takada2} models, 
the boundary values of $r_s$ exhibit significant 
differences between the two.  
One factor for the discrepancy should be that the present model is a 
2D system, while Ref.\cite{takada2} considers 3D systems.  
In the plasmon-mechanism analysis of 
superconductivity\cite{takada3} the critical $r_s$ in 2D is 
seen to be reduced (about 1/3) from that for 3D.  
More precisely, the present model differs from the Coulomb gas 
even though the 12-th neighbor interaction is included, 
since a cut-off in the interaction should degrade Gauss's law 
(hence degrade the plasmon modes).  
Extending our model to 3D may cast a light on this.

We thank Y. Takada for illuminating comments.
This work is in part supported by a Grant-in-Aid for Science Research on
Priority Area from the Japanese Ministry of Education.
Numerical calculations were performed at the supercomputer center, ISSP.

\bibliography{paper6}

\end{document}